\documentstyle[12pt]{article}

\begin{document}

\setcounter{page}{1}

\pagestyle{plain} \vspace{1cm}
\begin{center}
\Large{\bf Phantom Mimicry on the Normal Branch of a DGP-inspired Braneworld Scenario with Curvature Effect}\\
\small \vspace{1cm}  {\bf Kourosh Nozari$^{a,b}$}\quad and\quad {\bf Najmeh Alipour$^{a}$ }\\
\vspace{0.5cm} {\it $^{a}$Department of Physics, Faculty of Basic
Sciences,\\
University of Mazandaran,\\
P. O. Box 47416-95447, Babolsar, IRAN\\
\vspace{0.5cm} $^{b}$Research Institute for Astronomy and
Astrophysics of Maragha,
\\P. O. Box 55134-441, Maragha, IRAN\\
e-mail: knozari@umz.ac.ir}\\
\end{center}
\vspace{1.5cm}
\begin{abstract}
It has been shown recently that phantom-like effect can be realized
on the normal branch of the DGP setup without introduction of any
phantom matter neither in the bulk nor on the brane and therefore
without violation of the null energy condition. It has been shown
also that inclusion of the Gauss-Bonnet term in the bulk action
modifies this picture via curvature effects. Here, based on the
Lue-Starkman conjecture on the dynamical screening of the brane
cosmological constant in the DGP setup, we extend this proposal to a
general DGP-inspired $f(R,\phi)$ model that stringy effects in the
ultra-violet sector of the theory are taken into account by
inclusion of the Gauss-Bonnet term in the bulk action. We study
cosmological dynamics of this setup, especially its phantom-like
behavior and possible crossing of the phantom divide line especially
with a non-minimally coupled quintessence field on the brane. In
this setup, scalar field and curvature quintessence are treated in a
unified framework.
\end{abstract}
\newpage
\section{introduction}
Recent evidences from supernova searches data [1,2], cosmic
microwave background (CMB) results [3-5] and also Wilkinson
Microwave Anisotropy Probe (WMAP) data [6,7,8], show an positively
accelerating phase of the cosmic expansion today and this feature
shows that the simple picture of the universe consisting of the
pressureless fluid is not enough to describe the cosmological
dynamics. In this regard, the universe may contain some sort of the
additional negative-pressure dark energy. Analysis of the WMAP data
[9-11] shows that there is no indication for any significant
deviations from Gaussianity and adiabaticity of the CMB power
spectrum and therefore suggests that the universe is spatially flat
to within the limits of observational accuracy. Further, the
combined analysis of the WMAP data with the supernova Legacy survey
(SNLS) [9], constrains the equation of state $\omega_{de}$,
corresponding to almost $74\%$  contribution of dark energy in the
currently accelerating universe, to be very close to that of the
cosmological constant value. In this respect, a $\Lambda$CDM (
Cosmological constant plus Cold Dark Matter) model has maximum
agreement with the recent data, but with an unnaturally small and
fine-tuned value of $\Lambda$ [4]. From the general relativity
viewpoint, such models simply modify the right hand side of the
Einstein's field equations by including in the stress-energy tensor
something more than the usual matter and radiation components. As a
radically different approach, one can also try to leave the source
side unchanged, but rather modifying the geometric sector in the
left-hand side. In this viewpoint, the cosmic speed up can be
interpreted as a first signal of the breakdown of the laws of
physics as encoded by the standard General Relativity. This is
essentially reasonable: the General Relativity has been
experimentally tested only up to the Solar System scale and there is
no a priori theoretical motivation to extend its validity to
extraordinarily larger scales such as the cosmological ones.
Extending General Relativity without giving up its positive results,
opens the way to a large class of alternative theories of gravity
ranging from extra-dimensions [12] to non-minimally coupled scalar
fields [13,14,15]. In particular, fourth order theories [16,17,18]
based on replacing the scalar curvature $R$ in the Einstein-Hilbert
action with a generic analytic function $f(R)$ have attracted a lot
of attention in the last decade. Also referred to as $f(R)$ gravity,
these models have been shown to be able to both fit the cosmological
data and evade the Solar System constraints in several physically
interesting cases [19-28]. In this sense, modified gravity provides
the very natural gravitational alternative for dark energy. Also
modified gravity presents very natural unification of the early-time
inflation and late-time acceleration due to different role of
gravitational terms relevant at small and at large curvature
[29,30,31]. Moreover, some models of modified gravity are predicted
by string/M-theory considerations [16,17,18].

In the framework of braneworld scenarios, a possible approach to
realize the late-time acceleration of the universe is to consider
modification of gravity on the appropriate scales. This can be
achieved for instance by a weakening of gravitational interaction at
large scales. Weakening of gravity on large scales provides an
effective negative pressure that drives the late-time acceleration
of the universe. The Dvali, Gabadadze and Porrati (DGP) scenario
[32] is one of the simplest scenario in this regard. In this model,
our universe is a brane embedded in a Minkowski 5D bulk space-time.
Standard matter and gauge interactions are trapped on the brane and
only gravity experiences the full bulk. Cosmological dynamics in
this setup follows two distinct branches of solutions:
self-accelerating and the normal branch [33,34,35]. Late-time
acceleration can be explained in the self-accelerating branch
without invoking any unknown dark energy component. However, the
normal branch requires a dark energy component to accommodate the
current observations [36-45]. Although the self-accelerating branch
explains the late-time acceleration in a fascinating manner, it
suffers from serious theoretical problems such as the ghost issue
[46]. Fortunately the normal branch is free from instabilities.
Recently it has been shown that the normal branch of the DGP
braneworld scenario which is not self-accelerating can realize
phantom-like behavior without introducing any phantom field on the
brane [36-45]. By phantom-like behavior we mean an effective energy
density which is positive and grows with time and its equation of
state parameter stays always less than $-1$. Other extensions to
realize this phantom-like behavior are studied in Refs. [47-66]. On
the other hand, dark energy models with non-minimally coupled scalar
field and other extensions of scalar-tensor theories have been
studied widely some of which can be found in references [67-71].

In the spirit of scalar-tensor dark energy models, our motivation
here is to show that a general $f(R, \phi)$ GBIG-inspired scenario
can account for late-time acceleration and crossing of the phantom
divide line in some suitable domains of model parameters space. To
show this feature, first we study cosmological dynamics of $f(R,
\phi)$ GBIG-inspired scenario briefly. In the minimal case,
motivated by modified theories of gravity, some authors have
included a term of the type $l_{0}R^{n}$ in the action [72-77]. This
extension for some values of $n$ has the capability to explain
late-time acceleration of the universe in a simple manner. It is
then natural to extend this scenario to more general embeddings of
DGP inspired scenarios.

The purpose of this paper is to perform such a generalization to
study both late-time acceleration and possible crossing of the
phantom divide line in this setup. we construct a braneworld
scenario with modified induced gravity on the brane and the
Gauss-Bonnet term in the bulk action. We study cosmological dynamics
of this setup and we explore phantom-like behavior in the normal
branch of the model. We focus mainly on the potential of the type $V
(\phi) =\lambda\phi^{n}$ for non-minimally coupled scalar field. Our
setup for minimal case predicts a power-law acceleration supporting
observed late-time acceleration. In the non-minimal case, by a
suitable choice of non-minimal coupling and scalar field potential,
one obtains accelerated expansion in some specific regions of
parameters space. While a single minimally coupled scalar field in
four dimensions cannot reproduce a crossing of the phantom divide
line for any scalar field potential [47-66], a non-minimally coupled
scalar field account for such a crossing [78]. In GBIG model,
equation of state parameter of dark energy never crosses $\omega(z)
=-1$ line, and universe eventually turns out to be de Sitter phase.
Nevertheless, in this setup if we include a single scalar field
(ordinary or phantom) on the brane, we can show that equation of
state parameter of dark energy can cross the phantom divide line
[79]. Crossing of the phantom divide line with non-minimally coupled
scalar field on the warped DGP brane has been studied recently [80].
Here we present an extension of these non-minimal dark energy models
within GBIG-inspired $f(R, \phi)$ braneworld scenario.

We use a prime for differentiation with respect to $R$. An overdot
marks differentiation with respect to the brane time coordinate. We
show that this model provides a natural candidate for explanation of
several interesting aspects of the late-time acceleration without
suffering from issues such as the ghost instability. Among these
features, natural realization of the phantom divide line crossing is
worth to noting.

\section{The Modified GBIG Scenario}
The GBIG gravity is a generalized braneworld scenario which contains
both UV (ultra-violet) and IR (infra-red) limits in a unified
manner: It contains stringy effect via Gauss-Bonnet (GB) term in the
bulk action as the UV sector of the theory and Induced Gravity (IG)
effect which becomes important in the IR limit [81-89]. Here we add
a new ingredient to this scenario: we consider possible modification
of the induced gravity on the brane in the general framework of
$f(R,\phi)$ gravity. The action of our modified GBIG scenario is
written as follows
$$ S=\frac{1}{2\kappa_{5}^{2}}\int d^{5}x\sqrt{-g^{(5)}}\bigg\{R^{(5)}-
2\Lambda_{5}+\alpha\bigg[R^{(5)2}-4R_{ab}^{(5)}R^{(5)ab}+R_{abcd}^{(5)}R^{(5)abcd}\bigg]\bigg\}$$
\begin{equation}
+\int_{brane} d^{4} x
\sqrt{-g}\bigg[\frac{1}{\kappa_{5}^{2}}\overline{K}+\frac{1}{2\kappa_{4}^{2}}f(R,\phi)-\lambda+L_{m}\bigg]\
,
\end{equation}
where $\alpha$ is the GB coupling constant. We assume the brane is
located at $y=0$, where $y$ is the coordinate of the fifth
dimension. $\overline{K}$ is the trace of the mean extrinsic
curvature of the brane defined as follows
\begin{equation}
\overline{K}_{\mu\nu}=\frac{1}{2}\lim_{\epsilon\longrightarrow+0}
\Big([K_{\mu\nu}]_{y=-\epsilon}+[K_{\mu\nu}]_{y=+\epsilon}\Big)
\end{equation}
and its presence guarantees the correct matching conditions across
the brane. We denote the matter field Lagrangian by $L_{m}$  with
the following energy-momentum tensor
\begin{equation}
T_{\mu\nu}=-2\frac{\delta L_{m}}{\delta g^{\mu\nu}}+g_{\mu\nu}L_{m}
\end{equation}
The field equations resulting from the action (1) are given as
follows
\begin{equation}
M_{5}^{3}\Big(G_{B}^{A}+\alpha
H_{B}^{A}\Big)=\Lambda_{5}\delta_{B}^{A}+\delta_{\mu}^{A}\delta_{B}^{\nu}\tau_{\nu}^{\mu}.
\end{equation}
We note that a contribution of the form
$\frac{\sqrt{-g}}{\sqrt{-g^{(5)}}}$ is inherent in the definition of
$\tau_{\nu}^{\mu}$ on the right. The corrections to the Einstein
Field Equations originating in the GB term are represented by the
Lovelock tensor [90]
\begin{equation}
H_{B}^{A}=2RR_{B}^{A}-4R_{K}^{A}R_{B}^{K}-4R^{KL}R_{KBL}^{A}+2R^{AKLM}R_{BKLM}-\frac{1}{2}g_{B}^{A}L_{GB}.
\end{equation}
The energy-momentum tensor localized on the brane is
$$\tau_{\nu}^{\mu}=-M_{4}^{2}f'(R,\phi)G_{\nu}^{\mu}-\frac{M_{4}^{2}}{2}\Big[Rf'(R,\phi)-f(R,\phi)+
2\frac{\lambda}{M_{4}^{2}}\Big]\delta_{\nu}^{\mu}+T_{\nu}^{\mu}$$
\begin{equation}
+M_{4}^{2}[D^{\mu}D_{\nu}f'(R,\phi)-\delta_{\alpha}^{\mu}D^{\alpha}D_{\nu}f'(R,\phi)]
\end{equation}
where $D_{\mu}$ is the covariant derivative with respect to
$g_{\mu\nu}$ . The corresponding junction conditions relating
quantities on the brane are as follows [91,92]
$$\lim_{\epsilon\longrightarrow+0}[K_{\mu\nu}]_{y=-\epsilon}^{y=+\epsilon}=
\frac{f'(R,\phi)}{M_{5}^{3}}\bigg[\tau_{\mu\nu}-\frac{1}{3}g_{\mu\nu}g^{\alpha\beta}
\tau_{\alpha\beta}\bigg]_{y=0}-$$
\begin{eqnarray}
\frac{M_{4}^{2}}{M_{5}^{3}}f'(R,\phi)
\bigg[R_{\mu\nu}-\frac{1}{6}g_{\mu\nu}g^{\alpha\beta}R_{\alpha\beta}\bigg]_{y=0}.
\end{eqnarray}
To formulate cosmological dynamics on the brane, we assume the
following line element
\begin{equation}
ds^{2}=-n^{2}(y,t)dt^{2}+a^{2}(y,t)\gamma_{ij}dx^{i}dx^{j}+b^{2}(y,t)dy^{2}
\end{equation}
where $\gamma_{ij}$ is a maximally symmetric 3-dimensional metric
defined as $\gamma_{ij}=\delta_{ij}+k\frac{x_{i}x_{j}}{1-kr^{2}}$
where $k = -1, 0,+1$ parameterizes the spatial curvature and $r^{2}
= x_{i}x^{i}$. $y$ is the coordinate of extra dimension and the
brane is located at $y = 0$. The junction conditions on the brane
now gives the following expressions
$$\lim_{\epsilon\longrightarrow+0}[\partial_{y}n]_{y=-\epsilon}^{y=+\epsilon}(t)=
\frac{2n M_{4}^{2}}{M_{5}^{3}}\bigg[
\Big(\frac{df}{dR}\Big)\Big(\frac{\ddot{a}}{n^{2}a}-\frac{\dot{a}^{2}}{2n^{2}a^{2}}
-\frac{\dot{n}\dot{a}}{n^{3}a}-\frac{k}{2a^{2}}\Big)\bigg]_{y=0}$$
\begin{eqnarray}
+\frac
{n}{3M_{5}^{3}}\bigg[\Big(\frac{df}{dR}\Big)\Big(2\rho^{(tot)}+3p^{(tot)}\Big)
\bigg]_{y=0},\end{eqnarray}
\begin{eqnarray}\lim_{\epsilon\longrightarrow+0}[\partial_{y}a]_{y=-\epsilon}^{y=+\epsilon}(t)=
\frac{M_{4}^{2}}{M_{5}^{3}}\bigg[\Big(\frac{df}{dR}\Big)
\Big(\frac{\dot{a}^{2}}{n^{2}a}+\frac{k}{a}\Big)\bigg]_{y=0}-
\bigg[\bigg(\frac{df}{dR}\bigg)\frac{\rho^{(tot)}a}{3M_{5}^{3}}\bigg]_{y=0}.
\end{eqnarray}
If we choose a Gaussian normal coordinate system so that $b^{2}(y,
t) = 1$, these equations with non-vanishing components of the
Einsteins tensor in the bulk yield the following generalization of
the Friedmann equation for cosmological dynamics on the brane ( see
for instance [93-98] for the general machinery of this derivation )
 \begin{eqnarray}
 \bigg[1+\frac{8}{3}\alpha\Big(H^{2}+\frac{\Phi}{2}+\frac{K}{a^{2}}\Big)\bigg]^{2}
 \Big(H^{2}-\Phi+\frac{K}{a^{2}}\Big)=\bigg[r_{c}f'(R,\phi)H^{2}+r_{c}f'(R,\phi)\frac{K}{a^{2}}-
 \frac{\kappa^{2}_{5}}{6}\Big(\rho_{tot}+\lambda\Big)\bigg]^{2},
 \end{eqnarray}
where $\rho_{tot}\equiv\rho+\rho_{c}+\rho_{\phi}$. The bulk contains
a black hole mass and a cosmological constant so that $\Phi$ is
defined as follows
\begin{eqnarray}
\Phi+2\alpha\Phi^{2}=\frac{\Lambda_{5}}{6}+\frac{\Upsilon}{a^{4}}
\end{eqnarray}
where $\Upsilon$ is an integration constant. We note that $r_{c}$ is
the crossover scale in the DGP model which measures the strength of
the induced gravity effect on the brane and is related to the four
and five dimensional gravitational constants by
$r_{c}=\frac{\kappa^{2}_{5}}{2\kappa^{2}_{4}}$. Also, the parameter
$\alpha$ which is positive measures the strength of the GB curvature
effect on the brane and has dimension of length square. If
$\alpha=0$, the model reduces to the DGP model with modified induced
gravity, while for $r_{c}=0$ we recover the Gauss-Bonnet braneworld
model. Here we restrict our study to the case where the bulk black
hole mass vanishes (that is, $\Upsilon=0$) and therefore
$\Phi+2\alpha\Phi^{2}=\frac{\Lambda_{5}}{6}$. In this case the bulk
cosmological constant is given by
$\Lambda_{5}=\frac{-6}{l^{2}}+\frac{12\alpha}{l^{4}}$, where $l$ is
the bulk curvature. Assuming $\Lambda_{5}=0$ corresponding to a
Minkowski bulk, for a spatially flat FRW brane ($K=0$), the
Friedmann equation is given by
\begin{eqnarray}
\bigg[1+\frac{8}{3}\alpha\Big(H^{2}+\frac{\Phi}{2}\Big)\bigg]^{2}(H^{2}-\Phi)=\bigg[r_{c}f'(R,\phi)H^{2}
-\frac{\kappa^{2}_{5}}{6}\Big(\rho+\rho_{c}+\rho_{\phi}+\lambda\Big)\bigg]^{2},
\end{eqnarray}
where $\lambda$ is the brane tension. We restrict ourselves to the
normal branch of this DGP-inspired model. The energy density on the
brane is given by a CDM component ( with energy density $\rho_{m}$ )
and a cosmological constant $\Lambda$ with $\rho=\rho_{m}+\Lambda$
and a term $\rho_{c}$ originating on the modified induced gravity as
\begin{eqnarray}
\rho_{c}=\frac{1}{2}\bigg(R\frac{df(R,\phi)}{dR}-f(R,\phi)-6H\dot{R}\frac{d^{2}f(R,\phi)}{dR^{2}}\bigg)
\end{eqnarray}
\begin{eqnarray}
p^{(curv)}=\frac{M_{4}^{2}}{f'(R,\phi)}\bigg({2\dot{R}H
f''(R,\phi)+\ddot{R}f''(R,\phi)
+\dot{R}^{2}f'''(R,\phi)-\frac{1}{2}\Big[ f(R,\phi)-R
f'(R,\phi)\Big]}\bigg).
\end{eqnarray}
also
\begin{eqnarray}
\rho_{\phi}=\bigg[\frac{1}{2}\dot{\phi}^{2}+n^{2}V(\phi)-
6\frac{df}{d\phi}H\dot{\phi}\bigg]_{y=0},
\end{eqnarray}
\begin{eqnarray}p_{\phi}=\bigg[\frac{1}{2n^{2}}\dot{\phi}^{2}-V(\phi)+\frac{2}{n^{2}}
\frac{df}{d\phi}(\ddot{\phi}-\frac{\dot{n}}{n}\dot{\phi})+
4\frac{df}{d\phi}\frac{H}{n^{2}}\dot{\phi}+\frac{2}{n^{2}}
\frac{d^{2}f}{d\phi^{2}}\dot{\phi}^{2}\bigg]_{y=0},
\end{eqnarray}
where $\rho_{m}=\rho_{m_{0}}(1+z)^3$. With $\Lambda_{5}=0$ we find
$\Phi=0$ and $\Phi=-\frac{1}{2\alpha}$. Our forthcoming arguments
are based on the choice $\Phi=0$. Now, if we define the cosmological
parameters as\,
$\Omega_{m}=\frac{\kappa^{2}_{4}\rho_{m_{0}}}{3H^{2}_{0}}$,\,
$\Omega_{\Lambda}=\frac{\kappa^{2}_{4}\Lambda}{3H^{2}_{0}}$,\,
$\Omega_{r_{c}}=\frac{1}{4r_{c}^{2}H^{2}_{0}}$,\,\,
$\Omega_{\alpha}=\frac{8}{3}\alpha H^{2}_{0}$, \,
$\Omega_{c}=\frac{\rho_{c}\kappa^{2}_{4}}{3H^{2}_{0}}$ \,and \,
$\Omega_{\phi}=\frac{\rho_{\phi}\kappa^{2}_{4}}{3H^{2}_{0}}$,\, then
the Friedmann equation on the brane can be expressed as follows
\begin{eqnarray}
E^{2}(z)=\frac{1}{f'(R,\phi)}\Big[-2\sqrt{\Omega_{r_{c}}}E(z)\big[1+\Omega_{\alpha}E^{2}
(z)\big]+\Omega_{m}(1+z)^{3}+\Omega_{\phi}+\Omega_{c}+\Omega_{\Lambda}\Big]
\end{eqnarray}
where $E(z)=\frac{H}{H_{0}}$. Evaluating the Friedmann equation at
$z=0$ gives a constraint equation on the cosmological parameters of
the model as follows
\begin{eqnarray}
1+2\sqrt{\Omega_{r_{c}}}(1+\Omega_{\alpha})=\Omega_{m}+\Omega_{c}+\Omega_{\Lambda}+\Omega_{\phi}
\end{eqnarray}
which can be written formally as
\begin{eqnarray}
1+2\sqrt{\Omega_{r_{c}}}(1+\Omega_{\alpha})=\Omega_{m}+\Omega_{tot}.
\end{eqnarray}
We note that a contribution to this equation from
$f'(R,\phi)|_{z=0}$ has been normalized to unity. This is reasonable
since we finally have to choose an ansatz for $f(R,\phi)$ where $R$
and $\phi$ are related to the redshift via scale factor and its time
derivative. This quantity computed at $z=0$ gives a number which we
normalize it to unity. This constraint equation implies that
$\Omega_{m}+\Omega_{c}+\Omega_{\Lambda}+\Omega_{\phi}<1$ is
unphysical in the model parameter space. By imposing the condition
that the universe is currently accelerating, the deceleration
parameter defined as $q=-[\frac{\dot{H}}{H^{2}}+1]$ takes the
following form
\begin{eqnarray}
q=-\bigg[1-\frac{\dot{f'}(R,\phi)+\frac{3\Omega_{m}H_{0}(1+z)^{3}}{E(z)}-
\frac{\kappa^{2}_{4}(\dot{\rho_{c}}+\dot{\rho_{\phi}})}
{3H^{2}_{0}E^{2}(z)}}{H_{0}[2f'(R,\phi)E(z)+2\sqrt{\Omega_{r_{c}}}(1+\Omega_{\alpha}E^{2}(z))+
4\sqrt{\Omega_{r_{c}}}\Omega_{\alpha}E^{2}(z)]}\bigg].
\end{eqnarray}
To proceed further, we should specify the functional form of
$f(R,\phi)$. One of the most pressing problems of $f(R,\phi)$
theories is the need to escape the severe constraints imposed by the
Solar System tests [23],[99-101]. The Hu-Sawicki model with [25]
$$f(R)=R-R_{c}\frac{\alpha_{0}(\frac{R}{R_{c}})^{n}}{1+\beta_{0}(\frac{R}{R_{c}})^{n}}$$
is among those models that have good agreement with Solar-System
tests. In which follows, we choose the ansatz
\begin{eqnarray}
f(R,\phi)=\Big(R-R_{c}\frac{\alpha_{0}(\frac{R}{R_{c}})^{n}}{1+\beta_{0}(\frac{R}{R_{c}})^{n}}\Big)\Big(\frac{1}{2}(1-\xi
\phi^{2})\Big).
\end{eqnarray}
where the $R$-dependent part is from Hu-Sawicki scenario and in the
$\phi$-dependent part we have included a non-minimal coupling term.
We adopt also the ansatz $\phi(t)=\phi_{0}t^{-\mu}$ , where $\mu$ is
a positive constant and we set $\mu=0.9$. For scale factor we adopt
the following ansatz
\begin{equation}
a(t)=\Big(t^{2}+\frac{t_{0}}{1-\nu}\Big)^{\frac{1}{1-\nu}}
\end{equation}
where that $\nu\neq1$\, [102]. Figure $1$ shows the behavior of $q$
versus the redshift in this model. The universe enters the
accelerating phase at $z\approx 0.44$.
\begin{figure}[htp]
\begin{center}
\includegraphics{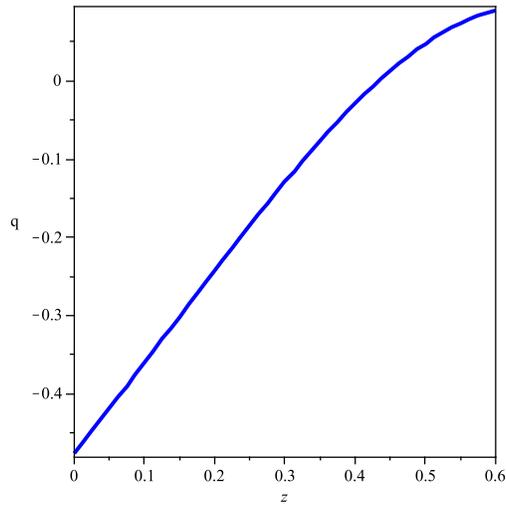}
\end{center}
\vspace{3 cm} \caption{\small {The variation of $q$ versus the
redshift.}}
\end{figure}

The modified Raychaudhuri equation follows easily from the Friedmann
equation of the brane and the conservation of the brane energy
density
\begin{eqnarray}
\frac{\dot{H}}{H^{2}_{0}}=-\frac{\dot{f'}(R,\phi)E^{2}(z)+3\Omega_{m}H_{0}(1+z)^{3}E(z)-
\frac{\kappa^{2}_{4}(\dot{\rho_{c}}+\dot{\rho_{\phi}})}
{3H^{2}_{0}}}{H_{0}[2f'(R,\phi)E(z)+2\sqrt{\Omega_{r_{c}}}
(1+\Omega_{\alpha}E^{2}(z))+4\sqrt{\Omega_{r_{c}}}\Omega_{\alpha}E^{2}(z)]}.
\end{eqnarray}
Figure $2$ shows the variation of $\frac{\dot{H}}{H^{2}_{0}}$ with
redshift. It is always negative and therefore there is no
super-acceleration and big-rip singularity in this braneworld
scenario. Since $\dot{H}<0$, the Hubble parameter decreases as the
brane expands.
\begin{figure}[htp]
\begin{center}
\includegraphics{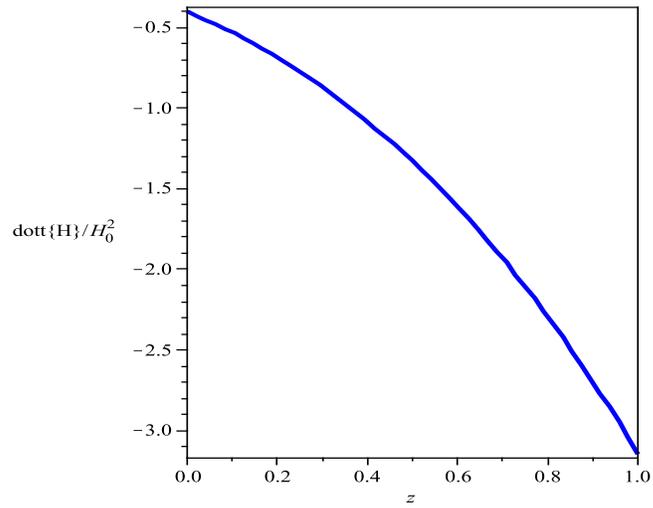}
\end{center}
\vspace{4 cm} \caption{\small { The variation of
$\frac{\dot{H}}{H^{2}_{0}}$ with redshift. }}
\end{figure}

\section{The Phantom-like Behavior}
We will show in  this section that a phantom-like behavior can be
realized on the brane and this occurs without including any exotic
matter that violates the null energy condition not on the brane nor
in the bulk. The phantom-like behavior is based on the definition of
an effective energy density which is corresponding to a balance
between the cosmological constant and geometrical effects encoded on
the Hubble rate evolution [36-45],[47-66]. The phantom-like behavior
on the brane is based on the definition of an effective energy
density $\rho_{eff}$ which increases with cosmic time and an
effective equation of state parameter always less than $-1$,\,
$\omega_{eff}<-1$. More precisely, the effective description is
inspired in writing down the modified Friedmann equation of the
brane as the usual relativistic Friedmann equation so that [47-66]
\begin{eqnarray}
H^{2}=\frac{\kappa^{2}_{4}}{3}(\rho_{m}+\rho_{eff}).
\end{eqnarray}
Using equations (13) and (25), we find
\begin{eqnarray}
\rho_{eff}=\frac{3H^{2}_{0}}{\kappa^{2}_{4}}\bigg[\frac{1}{f'(R,\phi)}\Big[-2\sqrt{\Omega_{r_{c}}}E(z)
(1+\Omega_{\alpha}E^{2}(z))+\Omega_{m}(1+z)^{3}+\Omega_{c}+
\Omega_{\phi}+\Omega_{\Lambda}\Big]-\Omega_{m}(1+z)^{3}\bigg].
\end{eqnarray}
By definition, phantom-like prescription breaks down if
$\rho_{eff}\leq0$. Figure $3$ shows the variation of $\rho_{eff}$
versus $z$ in our proposed setup.
\begin{figure}[htp]
 \begin{center}
 \includegraphics{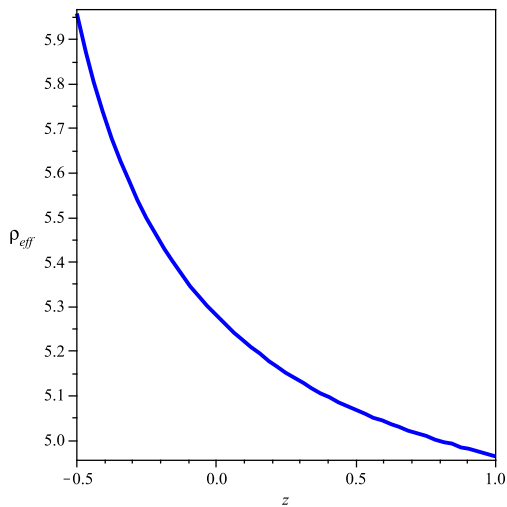}
 \end{center}
 \vspace{6 cm} \caption{\small { $\rho_{eff}$ versus redshift. }}
 \end{figure}
It is always positive and grows with cosmic time, therefore
effective energy density shows the phantom-like behavior. In this
respect, the important feature of this scenario is the fact that
this phantom-like behavior never breaks down at least in the
proposed subspace of the model parameter space. Nevertheless, there
are other subspaces of the model parameter space that phantom-like
prescription may be broken in these subspaces in the same way as has
been mentioned by Bouhamdi-Lopez in Ref. [47-66]. The main point
here is the fact that it is possible essentially to save
phantom-like prescription for a wide range of redshifts.

The effective equation of state parameter $\omega_{eff}$, can be
defined using the conservation equation of the effective energy
density on the brane
\begin{eqnarray}
\dot{\rho}_{eff}+3H(1+\omega_{eff})\rho_{eff}=0.
\end{eqnarray}
The effective energy density evolve as follows
$$\dot{\rho}_{eff}=\frac{-3}{\kappa^{2}_{4}}
\bigg[\frac{[2\dot{H}H_{0}\sqrt{\Omega_{r_{c}}}(1+\Omega_{\alpha}E^{2}(z))
+2\Omega_{\alpha}E^{2}(z)\dot{H}H_{0}+3\Omega_{m}E(z)(1+z)^{3}H^{3}_{0}-
\frac{\kappa^{2}_{4}\dot{\rho}_{c}+\dot{\rho}_{\phi}}{3}]}
{f'(R,\phi)}$$
\begin{eqnarray}
+\frac{\dot{f'}(R,\phi)H^{2}_{0}[-2\sqrt{\Omega_{r_{c}}}(1+\Omega_{\alpha}E^{2}(z))
+\Omega_{m}(1+z)^{3}+\Omega_{c}+\Omega_{\phi}+\Omega_{\Lambda}]}{(f'(R,\phi))^{2}}
-3\Omega_{m}(1+z)^{3}E(z)H^{3}_{0}\bigg].
\end{eqnarray}
By defining
\begin{eqnarray}
 y\equiv\frac{\dot{f'}(R,\phi)H^{2}_{0}[-2\sqrt{\Omega_{r_{c}}}(1+\Omega_{\alpha}E^{2}(z))
 +\Omega_{m}(1+z)^{3}+\Omega_{c}+\Omega_{\phi}+\Omega_{\Lambda}]}{(f'(R,\phi))^{2}}
 -3\Omega_{m}(1+z)^{3}E(z)H^{3}_{0},
 \end{eqnarray}
we find
\begin{eqnarray}
1+\omega_{eff}=\frac{\frac{-3}{\kappa^{2}_{4}}
[\frac{[2\dot{H}H_{0}\sqrt{\Omega_{r_{c}}}(1+\Omega_{\alpha}E^{2}(z))
+2\Omega_{\alpha}E^{2}(z)\dot{H}H_{0}+3\Omega_{m}E(z)(1+z)^{3}H^{3}_{0}-
\frac{\kappa^{2}_{4}\dot{\rho}_{c}+\dot{\rho}_{\phi}}{3}]}
{f'(R,\phi)}+y]}{3H_{0}E(z)\rho_{eff}}.
\end{eqnarray}
Figure $4$ shows the plot of $1+\omega_{eff}$ versus the redshift.
The universe enters to the phantom phase in $z\approx0.25$. As this
figure shows, the phantom-like behavior has no break down at small
redshifts in contrast to the $\Lambda$DGP-GB model [47-66]. So, our
model realizes a phantom-like behavior without introducing any
phantom matter on the brane or in the bulk. It is important to note
that this setup explains also a smooth transition to phantom phase,
the so called \emph{phantom-divide line crossing}.
\begin{figure}[htp]
\begin{center}
\includegraphics{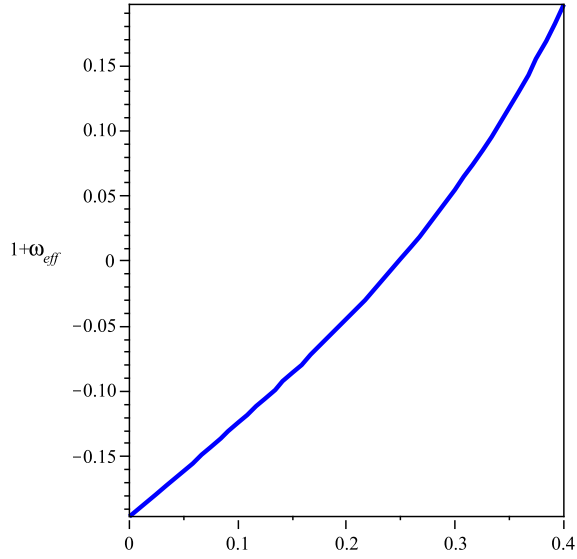}
\end{center}
\vspace{3.5 cm} \caption{\small { $1+w_{eff}$ versus the redshift.
}}
\end{figure}
Our main achievement here is the realization of the phantom-like
behavior without introducing any phantom matter not on the brane nor
in the bulk. Nevertheless, the phantom-like behavior realized in
this gravitational way should respect the null energy condition, at
least in some subspaces of the model parameter space.  So, we should
check the status of the null energy condition in this setup. The
null energy condition is respected if $\rho_{eff}+p_{eff}> 0$.
Figure $5$ shows a plot of $\rho_{eff}+p_{eff}$ versus the redshift.
As this figure shows, the phantom-like behavior in our proposed
model occurs without violation of the null energy condition at least
in some subspaces of the model parameter space. Especially, this
condition is respected in the phantom-like region of the model. We
emphasize that validity of the null energy condition in the
phantom-like region of the scenario is important for our purposes in
this paper. This is because with phantom fields usually the null
energy condition breaks down. Here we realized a phantom-like
behavior in a braneworld setup without violation of the null energy
condition in the phantom-like region. It is important to note that
in the non-phantom region, the situation is different and the
validity of null energy condition is guaranteed by positivity of the
effective energy density in this region.
\begin{figure}[htp]
\begin{center}
\includegraphics{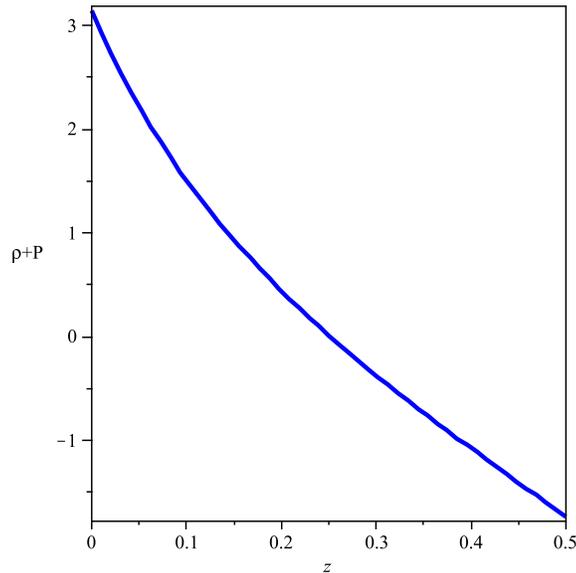}
\end{center}
\vspace{6 cm} \caption{\small { The status of the null energy
condition in this setup: $\rho_{eff}+p_{eff}$ versus the redshift.
}}
\end{figure}
\section{Lue-Starkman Screening of the Brane Cosmological Constant}
Lue and Starkman have shown that one can realize phantom-like
effect, that is, increasing of the effective dark energy density
with redshift, in the normal branch of the DGP cosmological solution
without introducing any phantom field [103]. The normal branch of
the DGP model which cannot explain the self-acceleration, has the
key property that brane is extrinsically curved so that shortcuts
through the bulk allow gravity to screen the effects of the brane
energy-momentum contents at Hubble parameters $H\sim r_{c}^{-1}$
where $r_{c}$ is the crossover distance [103]. Since in this case
$H(z)$ is a decreasing function of the redshift, the effective dark
energy component is increasing with redshift and therefore we
observe a phantom-like behavior without introducing any phantom
matter that violates the null energy condition and suffers from
several theoretical problems. In the first step, in this section we
study the phantom-like effect in the normal branch of a GBIG
scenario. In other words, here we suppose that there is a
quintessence field non-minimally coupled to modified induced gravity
on the brane in addition to a Gauss-Bonnet term in the bulk action.
In fact we incorporate the possibility of modification of the
induced gravity on the brane in the spirit of $f(R)$ gravity. We
emphasize also that we have included a canonical (quintessence)
scalar field to incorporate possible coupling of the gravity and
scalar degrees of freedom on the brane. This provides a wider
parameter space with capability to handle the problem more
completely. In fact, inclusion of this quintessence field brings the
theory to realize crossing of the phantom divide line [36-40].
Considering the normal branch of the equation (13) , we have
\begin{eqnarray}
H^{2}=\frac{8\pi
G}{3}(\rho_{m}+\rho_{\phi}+\rho_{c})+\frac{\Lambda}{3}-\frac{[f'(R,\phi)]^{-1}H}{r_{c}}(1+\frac{8}{3}\alpha
H^{2}),
\end{eqnarray}
where $G\equiv G_{eff}=\Big(8\pi M_{4}^{2}F'(R,\phi)\Big)^{-1}$ and
a prime denotes differentiation with respect to $R$. Now, we rewrite
this equation in the following form
\begin{equation}
H^{2}=\frac{8\pi G}{3}(\rho_{m}+\rho_{\phi}+\rho_{c})+\frac{8\pi
G}{3}(\rho_{DE}^{eff}).
\end{equation}
By comparing this equation with the Friedmann equation (31), we find
\begin{equation}
\frac{8\pi
G}{3}(\rho_{DE}^{eff})=\frac{\Lambda}{3}-\frac{[f'(R,\phi)]^{-1}H}{r_{c}}(1+\frac{8}{3}\alpha
H^{2}).
\end{equation}
As a result, in this GBIG-inspired model the crossover scale takes
the following form
\begin{equation}
r_{f}=\frac{M_{4}^{2}f'(R,\phi)}{2M_{5}^{3}}={f'(R,\phi)r_{c}},
\end{equation}
Now as an enlightening example, we set for instance
\begin{eqnarray}
f(R,\phi)=\Big(R-R_{c}\frac{\alpha_{0}(\frac{R}{R_{c}})^{n}}
{1+\beta_{0}(\frac{R}{R_{c}})^{n}}\Big)\Big(\frac{1}{2}(1-\xi
\phi^{2})\Big)\Big.
\end{eqnarray}
where the $f(R)$ part is the Hu-Sawicki model [25].  With this
choice, one recovers the general relativity for $n=0$. For $n\neq0$,
we obtain
\begin{equation}
\frac{8\pi
G}{3}(\rho_{DE}^{eff})=\frac{\Lambda}{3}-\frac{[\Big(R-R_{c}\frac{\alpha_{0}(\frac{R}{R_{c}})^{n}}
{1+\beta_{0}(\frac{R}{R_{c}})^{n}}\Big)\Big(\frac{1}{2}(1-\xi
\phi^{2})\Big)]^{-1}H}{r_{c}}(1+\frac{8}{3}\alpha H^{2})\,,
\end{equation}
where for a spatially flat FRW geometry, the Riici scalar is given
by
\begin{equation}
R=6\frac{\ddot{a}}{a}+6(\frac{\dot{a}}{a})^{2}.
\end{equation}
Figure $6$ shows the variation of $\rho_{eff}^{DE}$ versus $n$ and
the redshift in this GBIG-inspired $F(R,\phi)$ model. As we see, the
phantom-like behavior can be realized for all values of $n$
constraint in the Hu-Sawicki model. In other words, for all values
of acceptable $n$ in the Hu-Sawicki model, the effective dark energy
in this GBIG-inspired $F(R,\phi)$ model has the phantom-like
behavior.
\begin{figure}[htp]
\begin{center}\includegraphics{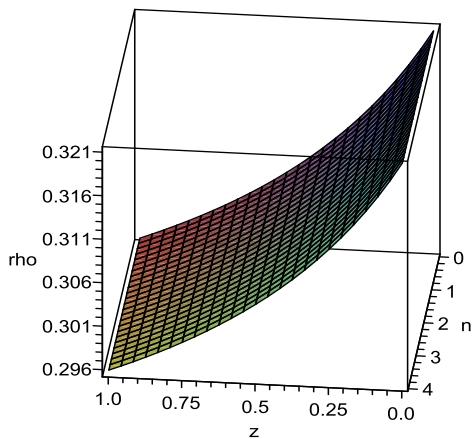}
 \vspace{6cm}
\end{center}
 \caption{\small {Variation of the effective dark energy density versus
$n$ and $z$ in a GBIG-inspired $F(R,\phi)$ model with
$F(R,\phi)=f(R,\phi)=\Big(R-R_{c}\frac{\alpha_{0}(\frac{R}{R_{c}})^{n}}
{1+\beta_{0}(\frac{R}{R_{c}})^{n}}\Big)\Big(\frac{1}{2}(1-\xi
\phi^{2})\Big)$. The phantom-like behavior can be realized for all
$n$.}}
\end{figure}
Figure $7$ shows the variation of the effective dark energy density
versus $z$ and the non-minimal coupling. By increasing the values of
$\xi$, the effective dark energy density reduces, and for a fixed
value of $\xi$ there is phantom-like effect for appropriate values
of $n$.
\begin{figure}[htp]
\begin{center}\includegraphics{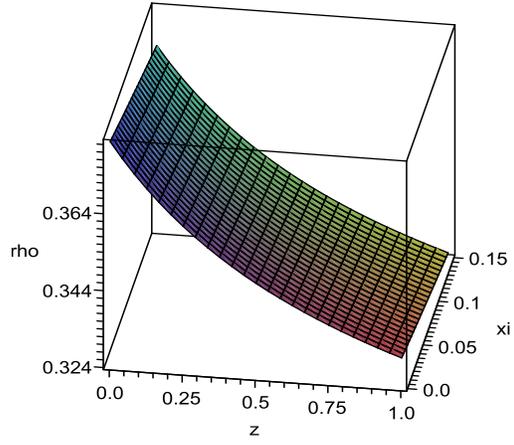} \vspace{4cm}
\end{center}
 \caption{\small {Variation of the effective dark energy density versus
$z$ and the non-minimal coupling.}}
\end{figure}
Also, figure $8$ shows the variation of the effective dark energy
density versus $n$ and the non-minimal coupling. The phantom-like
effect ( increasing the values of the effective dark energy density
with cosmic time) can be realized for suitable ranges of $n$ and
$\xi$.

\begin{figure}[htp]
\begin{center}\includegraphics{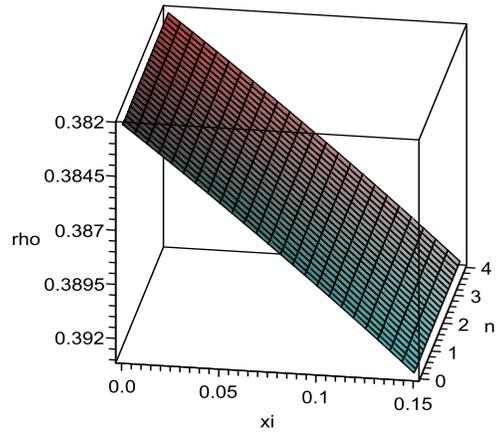} \vspace{4cm}
\end{center}
 \caption{\small {Variation of the effective dark energy density versus
$n$ and the non-minimal coupling.}}
\end{figure}
As an important especial case, for $F(R,\phi)=f(R)$ we find the
screening effect in a general GBIG $f(R)$-gravity
\begin{equation}
\frac{8\pi
G}{3}\rho_{DE}^{eff}=\lambda-\frac{\Big[f'(R)\Big]^{-1}H}{r_{c}}(1+\frac{8}{3}\alpha
H^{2}).
\end{equation}
Now as an enlightening example, we set for instance
\begin{equation}
f(R)=\Big(R-R_{c}\frac{\alpha_{0}(\frac{R}{R_{c}})^{n}}{1+\beta_{0}(\frac{R}{R_{c}})^{n}}\Big)
\end{equation}
Figure $9$ shows the variation of the effective dark energy density
versus $n$. There is phantom-like effect for all possible values of
$n$ in the Hu-Sawicki model.
\begin{figure}[htp]
\begin{center}
\includegraphics{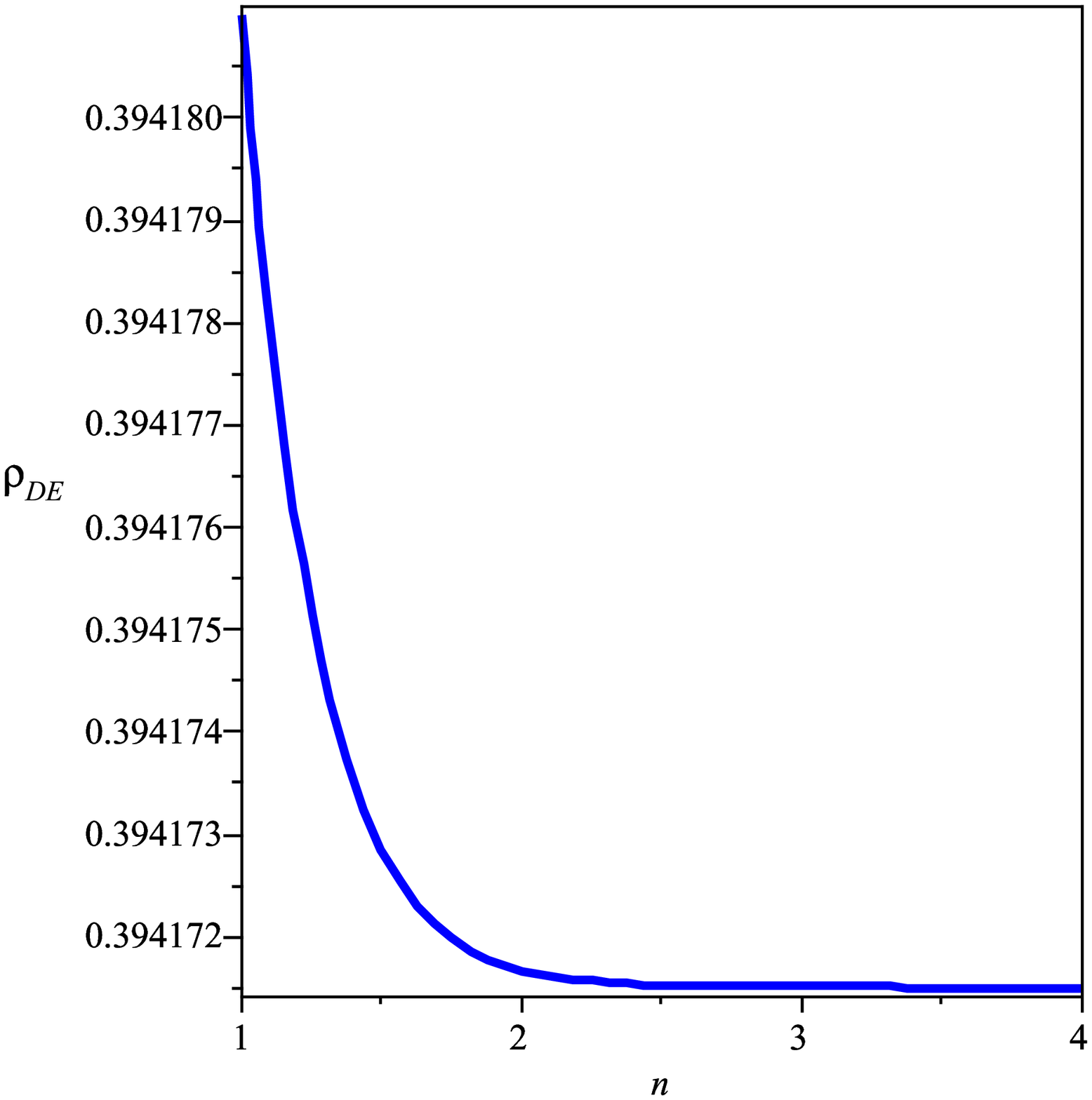}
\end{center}
\vspace{7 cm} \caption{\small { variation of the $\rho_{DE}$ versus
n. }}
\end{figure}
Figure $10$ shows the variation of the effective dark energy density
versus the redshift.
\begin{figure}[htp]
\begin{center}
\includegraphics{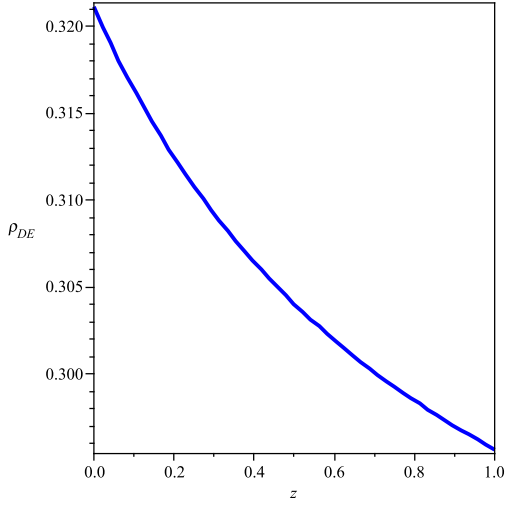}
\end{center}
\vspace{6.5 cm} \caption{\small { : $\rho_{DE}$ versus the redshift.
}}
\end{figure}
The main result of these numerical solutions is that this modified
GBIG scenario has the capability to realize the phantom-like
prescription without introducing any phantom field neither on the
brane nor in the bulk. We note that we have included a quintessence
field non-minimally coupled to the modified induced gravity on the
brane to have a wider parameter space and to include all possible
situations.

\section{Dynamical Screening and Phantom-Like Mimicry with
non-minimal coupling and curvature effect} Our setup in this section
contains a DGP-inspired braneworld scenario with curvature effect
through the Gauss-Bonnet term in the bulk action and a non-minimally
coupled scalar field on the brane. The ordinary matter localized on
the brane is assumed to be a quintessence scalar field non-minimally
coupled to the induced gravity on the brane. This canonical scalar
field respects the null energy condition and we assume there is no
phantom matter present on the brane or in the bulk. By virtue of
dynamical screening, we show that this model accounts for a
phantom-like behavior without violating the null energy condition.
In addition, we show that in contrast to existing literature which
argued that this phantom-like effect breaks down in some small
values of redshift, there is no break-down of this behavior in the
presence of the non-minimal coupling of the quintessence field . The
action of our model with Gauss-Bonnet and induced gravity effects
and a scalar field non-minimally coupled to induced gravity on the
brane is given as follows [104] (see also [81-84])
$$S=\frac{1}{2\kappa_{5}^{2}}\int d^{5}x\sqrt{-g^{(5)}}\Bigg[
R^{(5)}-2\Lambda_{5}+\alpha\Big(R^{(5)2}-4R_{ab}^{(5)}R^{(5)ab}+
R_{abcd}^{(5)}R^{(5)abcd}\Big)\Bigg]$$
\begin{equation}
+\Bigg[\frac{r_{c}}{2\kappa_{5}^{2}}\int
d^{4}x\sqrt{-g}\bigg(\beta(\phi) R -2\kappa_{4}^{2} g^{\mu\nu}
\nabla_{\mu}\phi\nabla_{\nu}\phi -4\kappa_{4}^{2}V(\phi) -4
\kappa_{4}^{2}\lambda\bigg)\Bigg]_{y=0},
\end{equation}
where $\alpha$ and $r_{c}$ are the GB coupling constant and IG
cross-over scale respectively. We assume the brane is located at
$y=0$ where $y$ is the coordinate of the extra dimension. Also
$\beta(\phi)$ is a general non-minimal coupling of the scalar field
and induced gravity on the brane. The cosmological dynamics of the
model is given by the following generalized Friedmann equation [104]
\begin{eqnarray}
\bigg[1+\frac{8}{3}\alpha\Big(H^{2}+\frac{\Phi}{2}+\frac{K}{a^{2}}\Big)\bigg]^{2}
\Big(H^{2}-\Phi+\frac{K}{a^{2}}\Big)=\bigg[r_{c}\beta(\phi)H^{2}+r_{c}\beta(\phi)\frac{K}{a^{2}}-
\frac{k^{2}_{5}}{6}\Big(\rho+\rho_{\phi}+\lambda\Big)\bigg]^{2}.
\end{eqnarray}
This equation describes the cosmological evolution on the brane with
tension and a non-minimally coupled scalar field localized on the
brane. As usual, the bulk contains a black hole mass and a
cosmological constant, so that $\Phi$ is defined as equation (12).
As usual, $r_{c}$ is the crossover scale in the DGP model and
$\alpha$ measures the strength of the GB curvature effect on the
brane. If $\alpha=0$, the model reduces to the DGP model with a
non-minimally coupled scalar field on the brane, while for $r_{c}=0$
we recover the Gauss-Bonnet braneworld model. Here, we restrict our
study to the case where the bulk black hole mass vanishes and we
neglect the bulk cosmological constant too ( that is,
$\Lambda_{5}=0$). For a spatially flat FRW brane, the Friedmann
equation is given by
\begin{eqnarray}
\bigg[1+\frac{8}{3}\alpha\Big(H^{2}+\frac{\Phi}{2}\Big)\bigg]^{2}(H^{2}-\Phi)=\bigg[r_{c}\beta(\phi)H^{2}
-\frac{\kappa^{2}_{5}}{6}\Big(\rho+\rho_{\phi}+\lambda\Big)\bigg]^{2}.
\end{eqnarray}
We restrict ourselves to the normal branch of this GBIG-inspired
scenario and therefore there is no ghost instabilities in this case.
The energy density on the brane $\rho $, is composed of a CDM
component ( with energy density $\rho_{m}$ ) and a cosmological
constant $\Lambda_{4}$
\begin{eqnarray}
\rho=\rho_{m}+\Lambda_{4}.
\end{eqnarray}
Considering the normal branch of the equation (43), we have
\begin{eqnarray}
H^{2}=\frac{8\pi
G}{3}(\rho_{m}+\rho_{\phi})+\frac{\Lambda_{4}}{3}-\frac{[\beta(\phi)]^{-1}H}{r_{c}}(1+\frac{8}{3}\alpha
H^{2}).
\end{eqnarray}
Comparing this equation with the following Friedmann equation
\begin{equation}
H^{2}=\frac{8\pi G}{3}(\rho_{m}+\rho_{\phi})+\frac{8\pi
G}{3}(\rho_{DE}^{eff})
\end{equation}
we find
\begin{equation}
\frac{8\pi
G}{3}(\rho_{DE}^{eff})=\frac{\Lambda_{4}}{3}-\frac{[\beta(\phi)]^{-1}H}{r_{c}}(1+\frac{8}{3}\alpha
H^{2})
\end{equation}
where $G\equiv G_{eff}=\Big(8\pi M_{5}^{2}\beta(\phi)\Big)^{-1}$ and
the crossover scale takes the following form
\begin{equation}
r_{f}=\frac{M_{4}^{2}\beta(\phi)}{2M_{5}^{3}}={\beta(\phi)r_{c}}.
\end{equation}
Equation (46) shows the dynamical screening of the brane
cosmological constant due to both non-minimal coupling and the
Gauss-Bonnet effect. In which follows, we set for simplicity
$\beta(\phi)=\frac{1}{2}(1-\xi \phi^{2})$ where $\xi$ is the
conformal coupling. We note also that in our numerical calculations
we adopt the ansatz $\phi(t)=\phi_{0}t^{-\mu}$ and
$a(t)=\Big(t^{2}+\frac{t_{0}}{1-\nu}\Big)^{\frac{1}{1-\nu}}$, where
$\mu$ and $\nu$ are positive constants. We obtain from equation (46)
\begin{equation}
\frac{8\pi
G}{3}(\rho_{DE}^{eff})=\frac{\Lambda_{4}}{3}-\frac{[\Big(\frac{1}{2}(1-\xi
\phi^{2})\Big)]^{-1}H}{r_{c}}(1+\frac{8}{3}\alpha H^{2}).
\end{equation}
Figure $11$ shows the variation of $\rho_{DE}^{eff}$ versus $\xi$ in
a constant time slice. The range of $\xi$ are chosen from [67-71]
constraint by the recent observations. As this figure shows, by
increasing the values of the nonminimal coupling, $\rho_{DE}^{eff} $
decreases in a fixed redshift( fixed time slice).
\begin{figure}[htp]
\begin{center}
\includegraphics{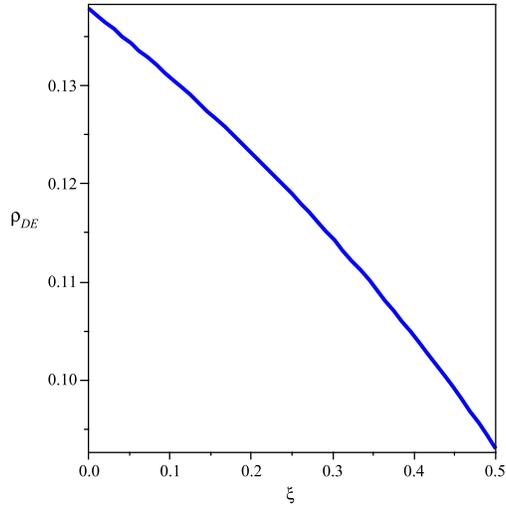}
\end{center}
\vspace{5 cm} \caption{\small { $\rho_{DE}^{eff}$ versus $\xi$ in a
constant time slice. }}
\end{figure}
Figure $12$ shows the variation of the effective dark energy density
versus the redshift. As this figure shows, $\rho_{DE}^{eff}$
increases by decreasing the redshift and this is exactly the
phantom-like behavior we are interested in. Note that this
phantom-like effects is realized without introducing any phantom
matter on the brane and only screening of the brane cosmological
constant due to non-minimal coupling and the curvature effect causes
such an intriguing aspect.
\begin{figure}[htp]
\begin{center}
\includegraphics{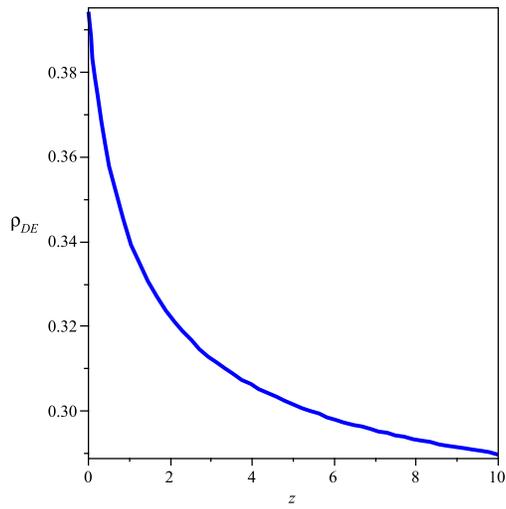}
\end{center}
\vspace{5 cm} \caption{\small { $\rho_{DE}$ versus redshift.  }}
\end{figure}

Figure $13$ shows the variation of the effective dark energy density
versus the redshift and the non-minimal coupling. We note that while
the introduction of a phantom field requires the violation of the
null energy condition, here this energy condition is respected ( at
least in some subspaces of the model parameter space) since we have
not included any phantom matter on the brane. In other words, since
the phantom-like dynamics realized in this setup is gravitational (
the quintessence field introduced on the brane plays the role of
standard matter on the brane), the null energy condition could be
respected in some subspaces of the model parameter space.
\begin{figure}[htp]
\begin{center}\includegraphics{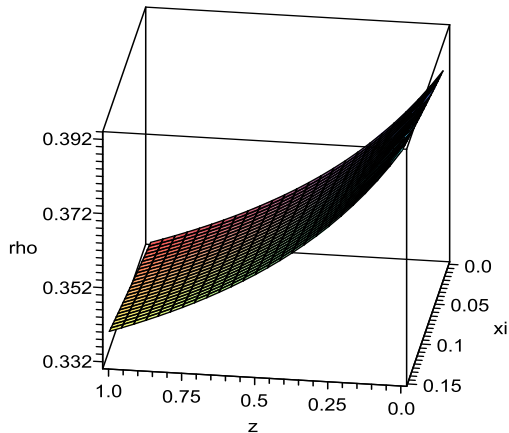} \vspace{7cm}
\end{center}
 \caption{\small {Variation of the effective dark energy versus the redshift and the non-minimal coupling.}}
\end{figure}
\section{Conclusion}
In this paper we have analyzed in some detail a modified
$\Lambda$DGP-Gauss-Bonnet model which corresponds to a 5D braneworld
model where the bulk is a 5D Minkowski space-time. This model
contains a Gauss-Bonnet term in the bulk and an induced gravity term
on the brane which is modified in the spirit of $f(R)$ gravity. We
incorporated also a canonical (quintessence) scalar field
non-minimally coupled to induced gravity on the brane. Our analysis
is performed for the normal (or non self-accelerating) ghost-free
branch which we have assumed to be filled by cold deark matter (CDM)
and a cosmological constant. We have shown especially that how the
brane accelerates at late-time in this normal branch. An interesting
feature of this model is the role played by the extra dimension: it
induces a mimicry of a phantom-like behavior without introducing any
matter that violates the null energy condition on the brane. This
phantom-like behavior happens without any super-acceleration of the
brane. We should emphasize that the Gauss-Bonnet effect which
induces the ultraviolet corrections on the brane, modifies the
phantom-like behavior at earlier times. Therefore, a consistent DGP
brane model which essentially is an infra-red modification of the
General Relativity, should have ultraviolet modifications as well
and this is achieved by incorporation of the Gauss-Bonnet term in
the bulk action. We have incorporated also the possible modification
of the induced gravity on the brane. This is equivalent to introduce
a canonical scalar field on the brane with General Relativity as the
background gravitational theory. By adopting suitable and
observationally viable ansatz for $f(R,\phi)$ and the scale factor,
we have shown that this model realizes the phantom-like behavior in
some subspaces of the model parameter space without violation of the
null energy condition. The phantom-like prescription in our proposed
model has no break down in some appropriate subspaces of the model
parameter space and it realizes a smooth crossing of the
phantom-divide line.\\

{\bf Acknowledgment}\\
This work has been supported partially by Research Institute for
Astronomy and Astrophysics of Maragha, IRAN.

\end{document}